\documentclass[aps,twocolumn,pra,tightenlines,floatfix,showpacs,superscriptaddress,longbibliography]{revtex4-2}
\usepackage{graphicx}
\usepackage{amsmath}
\usepackage{mathtools}
\usepackage{amssymb}
\usepackage{times}
\usepackage{epstopdf}
\usepackage[english]{babel}
\usepackage{natbib}
\usepackage{hyperref}
\hypersetup{colorlinks=true, citecolor=blue, linkcolor=blue}
\usepackage{color}
\usepackage{cleveref}
\usepackage{tablefootnote}
\usepackage{float}
\usepackage{subfig}
\usepackage{physics}
\usepackage{graphicx}
\usepackage{xcolor}
\usepackage{colortbl}

\usepackage{float}
\makeatletter
\let\newfloat\newfloat@ltx
\makeatother

\usepackage{algorithm}
\usepackage{algpseudocode}
\usepackage[version=4]{mhchem}

\DeclarePairedDelimiter{\ceil}{\lceil}{\rceil}

\begin{document}
\title{Compact Molecular Simulation on Quantum Computers via Combinatorial Mapping and Variational State Preparation}
\author{Diana Chamaki}
\affiliation{Applied Mathematics and Computational Research Division, Lawrence Berkeley National Laboratory, Berkeley, CA 94720, USA}
\affiliation{Department of Physics, University of California, Berkeley, Berkeley, CA 94720, USA}

\author{Mekena Metcalf}
\affiliation{Applied Mathematics and Computational Research Division, Lawrence Berkeley National Laboratory, Berkeley, CA 94720, USA}
\email{mmetcalf@lbl.gov}
\author{Wibe A. de Jong}
\affiliation{Applied Mathematics and Computational Research Division, Lawrence Berkeley National Laboratory, Berkeley, CA 94720, USA}

\begin{abstract}
Compact representations of fermionic Hamiltonians are necessary to perform calculations on quantum computers that lack error-correction. A fermionic system is typically defined within a subspace of fixed particle number and spin while unnecessary states are projected out of the Hilbert space. We provide a bijective mapping using combinatoric ranking to bijectively map fermion basis states to qubit basis states and express operators in the standard spin representation. We then evaluate compact mapping using the Variational Quantum Eigensolver (VQE) with the unitary coupled cluster singles and doubles excitations (UCCSD) ansatz in the compact representation. Compactness is beneficial when the orbital filling is well away from half, and we show at 30 spin orbital $\ce{H2}$ calculation with only 8 qubits. We find that the gate depth needed to prepare the compact wavefunction is not much greater than the full configuration space in practice. A notable observation regards the number of calls to the optimizer needed for the compact simulation compared to the full simulation. We find that the compact representation converges faster than the full representation using the ADAM optimizer in all cases. Our analysis demonstrates the effect of compact mapping in practice.
\end{abstract}
\maketitle
\section{Introduction}

Quantum computers promise to provide an advantage in chemistry and material simulation as molecular orbitals can be mapped to qubits~\cite{bassman2021review,McArdleReview2019,HM2022review}. This advantage is limited in the near term by the number of qubits and the quantity of operations required to capture correlation in the electronic wave function. Scaling the number of qubits on a quantum computer is an outstanding obstacle in the realization of error-corrected quantum computers, and even with error-correction the number of logical qubits will be limited. Without error correction, algorithm execution is limited by decoherence and noise, so the number of operations on a quantum computer needs to be greatly reduced. Hybrid quantum algorithms like the Variational Quantum Eigensolver (VQE) are used to prepare ground state wavefunction with shorter depth~\cite{mcclean2016theory}, realized in experiments performing elementary calculations of diatomic molecules in limited basis sets~\cite{Google2020Hartree,Kandala2017,Colless2018, OMalley2016}. Even with 100-1000 error-corrected qubits, second-quantized configuration interaction and couple cluster style chemistry simulations will be limited to small molecules with small basis sets. This means that reducing the number of qubits for atoms described with extended basis sets is essential to do meaningful chemistry simulations on quantum computers. 

A preliminary operation required for quantum simulation of fermions is mapping indistinguishable fermions basis to distinguishable spins. Widely used mapping schemes for second-quantized fermionic Hamiltonians bijectively map the fermion orbitals to qubits and field operators to spin operators~\cite{JW-1928,bravyi2002fermionic, McArdleReview2019, ParityMap}. These mapping techniques expand the full Hilbert space for a fermionic system without regard for symmetry and properties of the Hamiltonian. Accounting for symmetries reduces the number of configurations needed to express the wave function, leading to \textit{compact representations} of fermionic systems. There have been numerous approaches targeting compact Hamiltonian simulation in the matrix representation~\cite{moll2016optimizing, steudtner2018mapping,Bravyi2017tapering, Setia2021symm}. These approaches use spin operators to reduce the number of qubits to the relevant symmetry sector. Directly encoding fermion basis states to qubit basis states leads to the most compact representation of fermions on quantum computers, yet functional mapping forms are lacking in these works~\cite{Love-CI-2013,Shee2022}. Alternatively, one can encode only the occupied orbitals on the quantum computer and execute the algorithm using sparse Hamiltonian simulation~\cite{Kirby_Compact_2021}. 

We target compact quantum simulation of interacting fermion systems by representing Fock states and operators in a combinatoric representation that can be implemented on quantum computers. A combinatoric representation for fermion simulation is a natural choice to describe a Hilbert space with spin and number conserving properties. The combinatoric representation was recently introduced in Ref~\cite{Cederbaum_2010} as a matrix-free computational method to simulate many-body fermion and boson systems, and operations are reduced to simple algebra. This means the action of an operator on a state vector can be expressed using combinatoric algebra. Combinatoric simulation algorithms on quantum computers is an entirely new approach to fermion simulation that provides a framework for compact configuration interaction (CI) calculations on quantum computers. We employ near term tools like variational state-preparation and unitary matrix synthesis to create a compact representation of a second-quantized fermion system on a quantum computer, and we evaluate the performance of compact encoding in simulation. 

\section{Mapping Electronic Structure Hamiltonians to Quantum Computers} \label{Sec:el_structure}
We consider electronic structure calculations for molecules and materials on quantum computers using the second-quantized, molecular Hamiltonian,
\begin{align} \label{Eq:MolHam}
    &\hat{H} = \sum_{ij}\left(h_{ij}\hat{a}_i^\dagger \hat{a}_j + h_{ij}^*\hat{a}_j^\dagger \hat{a}_i \right)\nonumber\\
    &+ \frac{1}{2}\sum_{ijkl} \left[h_{ijkl}\left(\hat{a}_i^\dagger \hat{a}_j \hat{a}_k^\dagger \hat{a}_l - \delta_{kj}\hat{a}_i^\dagger \hat{a}_l\right)+{\rm H.C.}\right], 
\end{align}
where H.C. denotes the Hermitian conjugate.
The creation (annhilation) operators $\hat{a}_i^\dagger$ ($\hat{a}_j$) represent field operators on the molecular orbitals and obey anti-commutation relations,
\begin{align}
    \{\hat{a}_i, \hat{a}_j\} &= \{\hat{a}_i^\dagger, \hat{a}_j^\dagger\} = 0\\
    \{\hat{a}_i^\dagger, \hat{a}_j\} &= \delta_{ij}
\end{align}
to preserve anti-symmetry of the fermion wave function. The fermion basis is represented in the Fock basis,
\begin{align}
    |\psi\rangle &= \sum_i c_i |F_i\rangle\\
    |F\rangle &= \prod_{j}^N \hat{a}_j^\dagger |0\rangle
\end{align}
that creates a basis state containing N-particles by applying creation operators on the vacuum state $|0\rangle$. The Fock basis represents the occupancy, $n_i = (0,1)$, of the molecular orbitals. The Fock basis and the second-quantized operators are encoded onto the quantum computer to determine relevant observables.


Anti-symmetric wave functions for indistinguishable fermions must be mapped to distinguishable qubit basis states. A common mapping method for fermions – and most straightforward – is the Jordan Wigner transformation. The Jordan-Wigner transformation maps fermion Fock states of an M-orbital system directly to the quantum computational basis states of an M-qubit quantum register,

\begin{equation}
    |f_{M-1}, \, f_{M-2},\, ... \, f_{0}\rangle \rightarrow  |q_{M-1}, \, q_{M-2},\, ... \, q_{0}\rangle
\end{equation}
for $q_p = f_p \in {0,1}$ where $p \in [0,M)$.
The creation and annihilation operators are represented using Pauli spin operators,

\begin{align}
    \hat{a}_p = \frac{\hat{X} + i\hat{Y}}{2} \otimes \hat{Z}_{i-1} \otimes ... \otimes \hat{Z}_0 \\
    \hat{a}_p^\dagger = \frac{\hat{X} - i\hat{Y}}{2} \otimes \hat{Z}_{i-1} \otimes ... \otimes \hat{Z}_0, 
\end{align}
with  non-local $\hat{Z}$ operators account for parity and preserve the anti-commutation relations described above.
By transforming the Hamiltonian operators into the Pauli spin representation, the Hamiltonian is represented by a linear combination of the $\text{SU}(M)$ group,
\begin{equation}
    \hat{H} = \sum_{j=1}{c_j\hat{P}_j}
\end{equation}
where $c_j \in \mathbb{C}$ and $\hat{P}_j$ is a tensor product of Pauli operators. Fermionic mapping methods developed in this representation yield a Hamiltonian spanned by this group.

Widely used mapping schemes, such as Jordan-Wigner and Bravyi-Kitaev, result in a qubit operator which uses significantly more qubits than necessary. Techniques exist to remove qubits given symmetries in the system, and typically one qubit is removed for each symmetry~\cite{Bravyi2017tapering,Setia2021symm}. These qubit reduction techniques are beneficial for systems near half-filling. Significantly more qubits can be removed for systems that have occupations well away from half filling as the number of required basis states – considering fixed particle number and fixed spin – is $\binom{M}{N_{\uparrow}}\binom{M}{N_{\downarrow}}$ where M is the number of spatial orbitals, $N_{\uparrow}$ is the number of spin up particles, and $N_{\downarrow}$ is the number of spin down particles. Directly encoding fermionic basis states to qubit basis states increases the gate complexity of the Hamiltonian~\cite{moll2016optimizing,steudtner2018mapping}. Further advances have been made using majorana encodings and graphs resembling the toric code~\cite{derby2021mapping}. 

Here we focus on the maximum compressed mapping using a combinatorial ranking algorithm for variational algorithms and find advantages in convergence without high gate or measurement cost for molecular Hamiltonians. Recently, Ref. \cite{Shee2022} independently introduced an encoding scheme that is logarithmic in the number of qubits by mapping to basis states using an arbitrary map. This map lacked a functional form which we introduce using combinadics.  They performed VQE calculations using a hardware efficient ansatz, in contrast to the UCCSD ansatz we propose in the compact basis.

\section{Combinatoric Representation of Hamiltonians}
\label{Sec:Comb_Map}
Combinadics for matrix free fermionic and bosonic simulation has been proposed for classical computation~\cite{Cederbaum_2010}. We apply this set theoretic representation coupled with combinatoric algebra to map a fermionic basis states bijectively to qubit states and redefine unitary group generators as shift operations on the basis states. 
\subsection{Basis}

The set
\begin{equation} \label{full set}
    S = \{s \in \mathcal{N} : s < M\}
\end{equation}
is composed of natural numbers, $s$, that extend the orbital or lattice size $M$. Now we define a subset $\sigma \in S $ with cardinality N being the number of particles
\begin{equation} \label{sigma}
    \sigma = \{s_1,s_2,...s_N \}.
\end{equation}
 Each element $s_k$ represents an occupied orbital, and the integer $k$ is the position of the occupied orbital in the set. We can use these sets to represent the basis for a general wave function,
\begin{equation}
    \ket{\Psi} = \sum_i^{N_{conf}} c_i  \ket{\sigma_i}
\end{equation}
with fixed particle number, leaving the cardinality fixed. The dimension of the ordered set space scales by the number of fermion configurations $N_{conf} = \binom{M}{N}$ where $N$ particle positions are chosen from the base set of size $M$. 

The sets can be compactly mapped to qubit states by establishing a set-order. We order the sets lexicographically,
\begin{equation}
    s_1 < s_2 < ... < s_N, \forall \text{ }s \in \sigma.
\end{equation}
 Mapping the sets to unique natural numbers, known as combinadics, is performed under the assumption the each $\sigma$ is unique. The uniqueness of the lexicographically ordered sets enables a bijective mapping to a unique integer $I_\sigma$. We define a function 
\begin{equation} \label{f}
    f(|\sigma\rangle) = \binom{M}{N}-1 - \sum_{k=0}^{N-1} \binom{M-(s_{N-k}+1)}{k+1}.
\end{equation}
that operates on the elements of a set and yields a unique integer,
\begin{equation}
    |I_\sigma\rangle = f(|\sigma\rangle),
\end{equation}
\begin{equation}
    |\Psi\rangle = \sum_i^{N_{conf}}c_i |I_{\sigma,i}\rangle
\end{equation}
corresponding to the set. Using the function $f(\sigma)$ the natural numbers span a combinatoric Hilbert space
\begin{equation} \label{reduced set}
    \mathcal{I} = \{I_\sigma \in \mathcal{N} : I_\sigma < \binom{M}{N}\}
\end{equation}
which contains basis states in $\mathbb{C}^{2^M}$.

Extension to U(2M) composite Hilbert space $\mathcal{H}_\uparrow \otimes \mathcal{H}_\downarrow$ for fermions can be done two ways depending on how you order the spins. 1) An interleaved ordering, $|\downarrow \uparrow ... \downarrow \uparrow \rangle$, is trivially ranked lexicographically. 2) A block spin ordering $ | \downarrow \downarrow ...\downarrow\uparrow... \uparrow\uparrow\rangle$ can be mapped by taking the Cartesian product 
\begin{equation}
    \mathcal{I}^\uparrow \times \mathcal{I}^\downarrow = \{(I_\sigma^\uparrow, I_{\sigma^\prime}^{\downarrow}) : I_\sigma^\uparrow < \binom{M}{N_\uparrow} \text{ and } I_{\sigma^\prime}^\downarrow < \binom{M}{N_\downarrow}\}.
\end{equation}
Using a pairing function, we map the $\mathcal{N} \times \mathcal{N} \rightarrow \mathcal{N}$,
\begin{equation} \label{unique_integer}
I^{\uparrow \otimes \downarrow} = I_\sigma^\uparrow \binom{M}{N_\downarrow} + I_{\sigma^\prime}^\downarrow. 
\end{equation}
These integers, which can be trivially calculated, map bijectively to qubit basis states. Rather than mapping fermion orbitals to qubits, we map ranked fermion basis states to qubit computational basis states.

\subsection{Operators}
With the basis $\sigma$ established, one then turns to define Hamiltonian action on the basis states. We turn our focus to the action of number and spin preserving operations on the sets like the one-body fermionic operator $\hat{a}_i^\dagger \hat{a}_j$ that permutes a particle from orbital $j$ to orbital $i$. In the compact representation the operations on the sets are no longer permutations. This operator, known as the group generator, acts on the sets as,
\begin{align}
    \hat{a}_i^\dagger \hat{a}_j |\sigma\rangle &= 
    \begin{cases}
    (-1)^p|(\sigma/\{j\})\cup \{i\}\rangle& ;  i \notin \sigma, \text{  }j \in \sigma \\
    0 & \text{otherwise}
    \end{cases}\\
    p &= |\{x\}: x\in \sigma \cap (\min(i,j),\max(i,j))| .
\end{align}
We are interested in the form of these operations after applying the function $f(\sigma)$ to the sets. The exclusion and inclusion operations on the sets transform into addition and subtraction operations on the function. We define an operation, $\hat{E}_{ij} \equiv \hat{a}_i^\dagger \hat{a}_j$ on the ranked sets that acts as a shift on the combinatorial number line,
\begin{align}
    \hat{E}_{ij} &= \sum_m^{N_{conf}}\sum_n^{N_{conf}}  |I_m\rangle\langle I_m|h(i) \circ g(j)|I_n\rangle \langle I_n|\\
    &= \sum_m^{N_{conf}}\sum_n^{N_{conf}} c_{mn} |I_m\rangle\langle I_n|\\
    g(j) &= \binom{M-(j+1)}{r+1}\\
    h(i) &= -\binom{M-(i+1)}{q+1}.\\
    c_{mn} &= 
    \begin{cases}
    (-1)^{|q-r|} & i \notin \sigma, \text{  }j \in \sigma\\
    0 & \text{otherwise}
    \end{cases}
\end{align}
where $s_q = i$ and $s_r = j$ are elements in the set $\sigma$. Parity of the fermionic operations is preserved by considering the position of the elements, $p = |q-r|$. The operator can be expressed in simplified equation,
\begin{align}
    E_{ij} &= \sum_n^{N_{conf}} c_{n} |I_n + d(i,j)\rangle\langle I_n|\\
    d(i,j) &= h(i) \circ g(j)\\
    d(i,j) &\in \mathcal{Z}.
\end{align}
We see that the action of the group generator connects numbers on the combinatorial number line using addition and subtraction. All the numbers, i.e. basis states, are connected under these operations. 
Furthermore, the operators obey the commutation relation
\begin{align}
    \left[\hat{E}_{ij},\hat{E}_{kl}\right] &= \delta_{jk}\hat{E}_{il} - \delta_{li}\hat{E}_{kj}.
\end{align}
as subtraction is a non-commutative operation. 

Since our fermionic Hamiltonian is non-linear we need to evaluate how the Coulomb interaction, two-body operators, are expressed on the number line. The two-body interaction is a composite of the unitary group generators,
\begin{equation}
    \hat{E}_{ij}\hat{E}_{kl} \equiv \hat{a}_i^\dagger \hat{a}_j \hat{a}_k^\dagger \hat{a}_l. 
\end{equation}
We can expand the operator as we did for the one-body operator and use the technique to simplify the equation,
\begin{widetext}
\begin{align}
    \hat{E}_{ij}\hat{E}_{kl} &= \sum_{mnop}^{N_{conf}} |I_m\rangle\langle I_m|h(i) \circ g(j)|I_n\rangle \langle I_n|I_o\rangle\langle I_o|h(k) \circ g(l)|I_p\rangle \langle I_p|\\
    &= \sum_{mnop}^{N_{conf}}c_{op}c_{mn} |I_m\rangle\langle I_n|I_o\rangle\langle I_p|\\
    &= \sum_{mnop}^{N_{conf}}c_{op}c_{mn} |I_n + d(i,j)\rangle\langle I_n|I_p + d(k,l)\rangle\langle I_p|\\
    &= \sum_{mnop}^{N_{conf}} c_{op}c_{mn} |I_p+d(k,l) +d(i,j)\rangle \langle I_p|.
\end{align}
\end{widetext}
Given quantum orthogonality conditions, we see that the two-body operator acts as two shifts on the basis state, and is comprised of a similar structure as the one-body operator.

\subsection{Hamiltonian Formulation}
In order to apply our quantum representation to quantum chemistry problems we need to switch representations for the molecular Hamiltonian in Eq(\ref{Eq:MolHam}). The action of the shift operators is equivalent to the action of a one-body operation on a fermion particle. Graph methods of for classical, configuration interaction calculations refer to the one-body operator $\hat{a}_i^\dagger \hat{a}_j$ as the generator of fermionic action and is equivalent to the shift operator \cite{DuchNotes}, therefore,
\begin{equation}
    \hat{E}_{ij} \equiv \hat{a}_i^\dagger \hat{a}_j.
\end{equation}
Using this definition and the properties of shift operators, we can easily re-write the molecular Hamiltonian entirely from our shift operators,
\begin{widetext}
\begin{equation}\label{Eq:ShiftHam}
    \hat{H} = \sum_{ij}\left(h_{ij}\hat{E}_{ij} + h_{ij}^*\hat{E}_{ij}^\dagger\right) + \sum_{ijkl} \left[h_{ijkl}\left(\hat{E}_{ij}\hat{E}_{kl} + \delta_{kj}\hat{E}_{il}\right)+h^*_{ijkl}\left(\hat{E}_{kl}^\dagger E_{ij}^\dagger + \delta_{kj}\hat{E}_{il}^\dagger\right)\right].
\end{equation}
\end{widetext}
The molecular Hamiltonian in the combinatoric representation has the form of Eq. (\ref{Eq:ShiftHam}) which we can implement on a quantum computer.

\section{Applications for NISQ devices} \label{Sec:NISQ}

The combinatoric mapping scheme described in Section~\ref{Sec:Comb_Map} can be applied to fermionic Hamiltonian simulation. We implement the mapping to build a  molecular Hamiltonian with a reduced number of qubits, $n$. Our compact mapping scheme is motivated by the limited number of qubits available on quantum hardware lacking error corrected schemes. Therefore we evaluate compactness with near-term variational state-preparation approaches. We use the hybrid classical-quantum Variational Quantum Eigensolver (VQE) algorithm to find the ground state energy of compact molecular Hamiltonians. VQE is based on the Rayleigh-Ritz variational principle, which ensures that the energy expectation value of a parameterized wave function will always be greater than or equal to the ground state energy,
\begin{equation}
    \langle \psi(\theta) \vert \hat{H} \vert \psi(\theta) \rangle  \geq E_0.
\end{equation}
 In our implementation of VQE, we use the compact Hamiltonian and define a UCCSD ansatz in the combinatoric representation. 

\subsection{Simulating the Molecular Hamiltonian}\label{mol_ham}

Current approaches to applying the Hamiltonian on parameterized wave functions require expressing the Hamiltonian in the Pauli group on n qubits to determine the appropriate basis for energy measurement. There exist novel approaches for fault-tolerant computers using a linear combination of unitaries and oracular Hamiltonian encoding, which our combinatoric approach is amenable to. However, we focus on the Pauli group approach in our following evaluation~\cite{Kirby_Compact_2021}. We express the compact Hamiltonian, $\hat{H}_c$, within the Pauli group and evaluate the combinatoric mapping showcased in the construction of the matrix. 

We express the Hamiltonian in matrix form using algorithm~\ref{alg:cap} and write it as a linear combination of tensor products of Pauli matrices
\begin{align}\label{qubit_ham_rep}
    \hat{H}_c &= \sum_{j=1}^m{c_j\prod_{i}^{n}\hat{\sigma}_{i}}\\
    &= \sum_{j=1}^m{c_j\hat{P}_j}, \nonumber
\end{align}
where $m = 4^n$ and $n=\ceil{\log _{2}(\binom{M}{N_{\uparrow}}\binom{M}{N_{\downarrow}})}$. We are able represent the qubit Hamiltonian as shown in Eq. \eqref{qubit_ham_rep} because elements of the Pauli group on n qubits equipped with the Hilbert Schmidt inner product form a basis for $M_{2^n}(\mathbb{C})$, where $M_{2^n}(\mathbb{C})$ is the Hilbert space of all $2^n \times 2^n$ matrices equipped with the Hilbert Schmidt inner product. The Hilbert Schmidt inner product is defined as follows: if $\hat{A}, \hat{B}$ $\in M_{2^n}(\mathbb{C})$
\begin{equation}
    \langle \hat{A},\hat{B}\rangle_{HS} = Tr(\hat{A}^*\hat{B})
\end{equation}
Then, we use the Hilbert Schmidt norm given by Eq. \eqref{hs_norm} to normalize the Hamiltonian.
\begin{align} \label{hs_norm}
    \langle \hat{A},\hat{A}\rangle_{HS} &= Tr(\hat{A}^*\hat{A})\\
    &=(\sum_{i=1}^n \sum_{j=1}^n a_{i,j})^{1/2} \nonumber
\end{align}
where $a_{i,j}$ are the matrix elements of $\hat{A}$.



Normalizing the basis yields a factor of $\frac{1}{\sqrt{2^n}}$ for each of the terms in Pauli group on n qubits. Then, to calculate coefficients in \eqref{pauli_decomp} we use the Hilbert Schmidt inner product,
\begin{equation} \label{coefficient_calc}
    c_j = \langle\,\hat{H}_c,\hat{P}_j\rangle_{HS}.
\end{equation}
Thus, the final form of the decomposed Hamiltonian is
\begin{equation} \label{pauli_decomp}
    \hat{H} = \frac{1}{\sqrt{2^n}}\sum_{j=1}^m{c_j\hat{P}_j}.
\end{equation}
 Table \ref{pauli_terms_chart} in section \ref{results} quantitatively illustrates the number of non zero coefficients in Eq. \eqref{pauli_decomp} for $\ce{H2}$. Once we have the compactly encoded Hamiltonian in its final decomposed form, we pass it into VQE to calculate the ground state energy. 

\subsection{Compactly Encoded, Chemically Inspired Anzatz}\label{ansatz}

Before we can run VQE with the compact Hamiltonian we need an ansatz to prepare the wave function in a compact basis. We construct a UCCSD ansatz in this compact representation for the VQE algortithm. Our UCCSD operator rotates the wave function on the axis of single excitations and double-excitations to capture correlation.  The UCCSD is operator is 
\begin{align}
    U(\vec{\theta}) &= e^{\hat{T}-\hat{T}^\dagger}\\
    \hat{T} &= \hat{T}_1 + \hat{T}_2\\
    \hat{T}_1 &= \sum_{\alpha\in virt, i \in occ} \theta_i^\alpha\hat{a}_\alpha^\dagger \hat{a}_i\\
    \hat{T}_2 &= \sum_{\alpha \beta\in virt, ij \in occ} \theta_{ij}^{\alpha \beta}\hat{a}_\alpha^\dagger \hat{a}_\beta^\dagger \hat{a}_i \hat{a}_j 
\end{align}
where occ are the occupied orbitals denoted with Latin characters and virt are the virutal orbitals denoted by Greek characters \cite{mcardle2020quantum}. The operator is applied to a reference state $|\phi_0\rangle$ and the amplitudes $t_{i\alpha}$ and $t_{ij\alpha \beta}$ are parameterized excitation amplitudes determined variationally. The Hartree-Fock ground state is typically used as the reference function. In the combinatoric representation the Hartree Fock ground state is the computational basis state, so the parameterized wave function becomes
\begin{equation}
    |\Psi(\theta)\rangle = U(\theta)|0\rangle.
\end{equation}
The operator $U(\theta)$ needs to be re-expressed in the combinatoric representation leading to an interesting unitary structure. The UCCSD operator rotates the wave function from the reference function into a Hilbert space containing singly and double excited Slater determinants, so we re-express the excitation operators as
\begin{align}
        \hat{T}_1 &= \sum_{i\in virt, \alpha \in occ} \theta_{i}^{\alpha}|\psi^\alpha_i\rangle\langle 0| + \text{H.C.}\\
    {T}_2 &= \sum_{ij\in virt,\alpha \beta \in occ} \theta_{ij}^{\alpha\beta}|\psi^{\alpha\beta}_{ij}\rangle\langle 0| + \text{H.C.},
\end{align}
where $|\psi_\alpha^i\rangle$ is a single excited determinant and $|\psi_{\alpha\beta}^{ij}\rangle$ is double excited determinant.

We express the UCCSD operator as a time-evolution operation $\hat{U} = e^{-i\hat{K}}$ with $\hat{K} = i(\hat{T}_c-\hat{T}_c^\dagger)$ to implement on a quantum computer  ~\cite{McCleanThesis}.  The excitation operator $\hat{K}$ is a rank-2 matrix of the same dimensionality as the Hamiltonian. This means our cluster operator contains two singular values and the rest of the values are exactly 0. $\hat{T}_c$ the subspace compressed cluster operator we developed. It is populated by zeros everywhere except the first row. The first row is populated with parameters, which are placed based according to the excitation index for single and double excitations. For example, if we had a system that required n qubits using our mapping scheme with parameters $\theta_i$ we would have the following matrix, 
\begin{equation}
\hat{K} = i(\hat{T}_c - \hat{T}_c^\dagger) = \begin{bmatrix}
0 & i\theta_1 &  \dots & i\theta_n\\
-i\theta_1 & 0&\dots & 0\\
\vdots & \vdots & \ddots & \vdots\\
-i\theta_n & 0 & \dots & 0
\end{bmatrix}
\end{equation}

We decompose $\hat{K}$ in the same manner that we decomposed the Hamiltonian using the Hilbert Schmidt norm to get a linear product of Pauli operators and construct an evolution circuit.

\subsection{Results and Discussion} \label{results}

\begin{table*}[ht]
{\renewcommand{\arraystretch}{1.17}{%
\begin{tabular}{c@{\hspace{0.5cm}}c@{\hspace{0.5cm}}c@{\hspace{0.5cm}}c@{\hspace{0.5cm}}c@{\hspace{0.5cm}}c@{\hspace{0.5cm}}c}
\hline
\begin{tabular}[c]{@{}l@{}}Bond Distances\end{tabular} &
\begin{tabular}[c]{@{}l@{}l@{}}4 Spin Orbitals\\ on 2 Qubits\end{tabular} & \begin{tabular}[c]{@{}l@{}l@{}}8 Spin Orbitals \\ on 4 Qubits\end{tabular} & \begin{tabular}[c]{@{}l@{}l@{}}10 Spin Orbitals \\ on 5 Qubits\end{tabular} & \begin{tabular}[c]{@{}l@{}l@{}}16 Spin Orbitals \\ on 6 Qubits\end{tabular} &
\begin{tabular}[c]{@{}l@{}l@{}}22 Spin Orbitals \\ on 7 Qubits\end{tabular} &
\begin{tabular}[c]{@{}l@{}l@{}}30 Spin Orbitals \\ on 8 Qubits\end{tabular} \\\hline
0.35 \AA& 8& 158& 752& 2042& 16384&62354\\\arrayrulecolor{lightgray}\hline
0.55\AA& 8& 128& 512& 3699& 16381&60827\\\hline
0.7\AA& 8& 123& 512& 3647& 16382&62804\\\hline
1.4\AA& 8& 125& 510& 3649& 16314&61719\\\hline
2.8\AA& 5& 125& 508& 3488& 16197&63119\\\hline
4.4\AA& 7& 213& 1012& 3743& 16181&63855\\\hline
6.0\AA& 13& 252& 1013& 4035& 16168&64782\\ \arrayrulecolor{black}\hline
\end{tabular}}
}\caption{The number of Pauli operators in the compactly encoded $\ce{H2}$ Hamiltonian for each bond distance using the decomposition method described in section \ref{mol_ham}.}\label{pauli_terms_chart}
\end{table*}

\begin{figure}[H] 
    \centering
    \includegraphics[width=\linewidth]{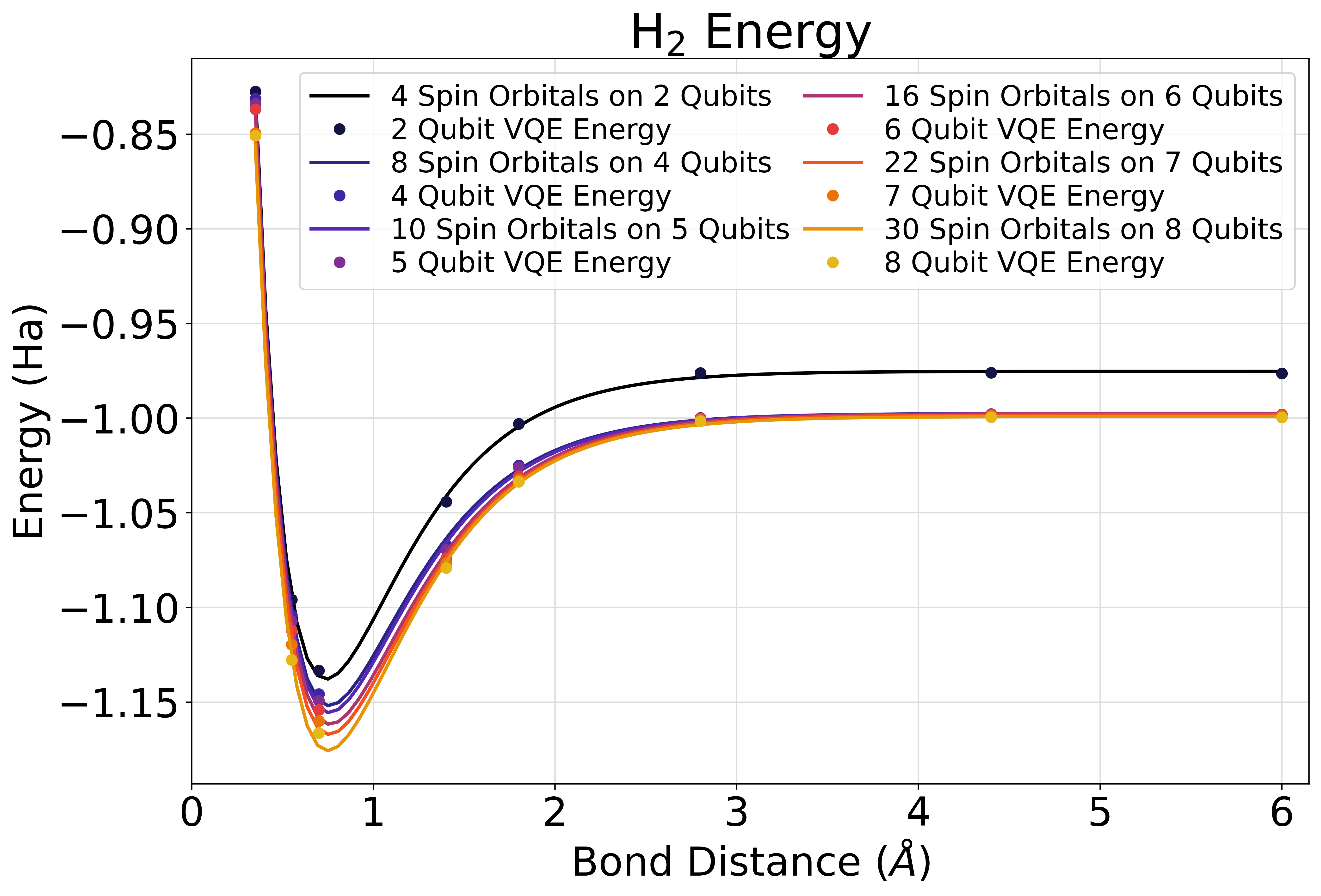}
    \caption{The exact ground state energies of $\ce{H2}$ at the given active spin orbitals as a function of bond distance. We use the Morse potential to create the curve fit for the exact energies obtained through exact diagonalization. Each point represents the VQE energy obtained using the compactly encoded Hamiltonian and UCCSD-inspired ansatz.}
    \label{fig:h2_energy} 
\end{figure}

We implemented the compactly encoded Hamiltonian and UCCSD-inspired ansatz described in sections \ref{mol_ham} and \ref{ansatz} for $\ce{H2}$ and $\ce{LiH}$ in the cc-pVTZ basis using the IBM qasm simulator. Fig. \ref{fig:h2_energy} shows the results for $\ce{H2}$ with 4, 8, 10, 16, 22, and 30 active spin orbitals each mapped to 2, 4, 5, 6, 7, 8 qubits respectively. This reduction enables us to obtain energies for a once 30 qubit calculation using only 8 qubits. 

To ensure the accuracy of the VQE energies, we calculated the exact ground state energies by diagonalizing the compactly encoded Hamiltonian to get the exact ground state energy. Then, for $\ce{H2}$ with 4, 8 active orbitals, we checked the results from the compactly encoded Hamiltonian diagonalization to the results from diagonalizing the full Hamiltonian with the Jordan-Wigner mapping. We successfully obtained the VQE ground state energies of $\ce{H2}$ within $1.6 \times 10^{-3}$ Ha of the exact values using the gradient based ADAM optimizer Fig.~\ref{fig:h2_energy}. Mappings with bijective correspondence between fermion orbitals and qubits require as many qubits as spin orbitals; thus, the 30 active spin orbital calculation would have required 30 qubits. The ground state energy of $\ce{H2}$ with 30 active spin orbitals is calculated within chemical accuracy using compact mapping on 8 qubits. The number of Pauli operators in the Hamiltonian given in Eq. \eqref{pauli_decomp} for $\ce{H2}$ using 4, 8, 10, 16, 22, and 30 active spin orbitals is presented in Table \ref{pauli_terms_chart}. The varying number of terms in the Hamiltonian results from the number of integrals needed to describe the molecular system.

\begin{figure}[H]
    \centering
    \includegraphics[width=\linewidth]{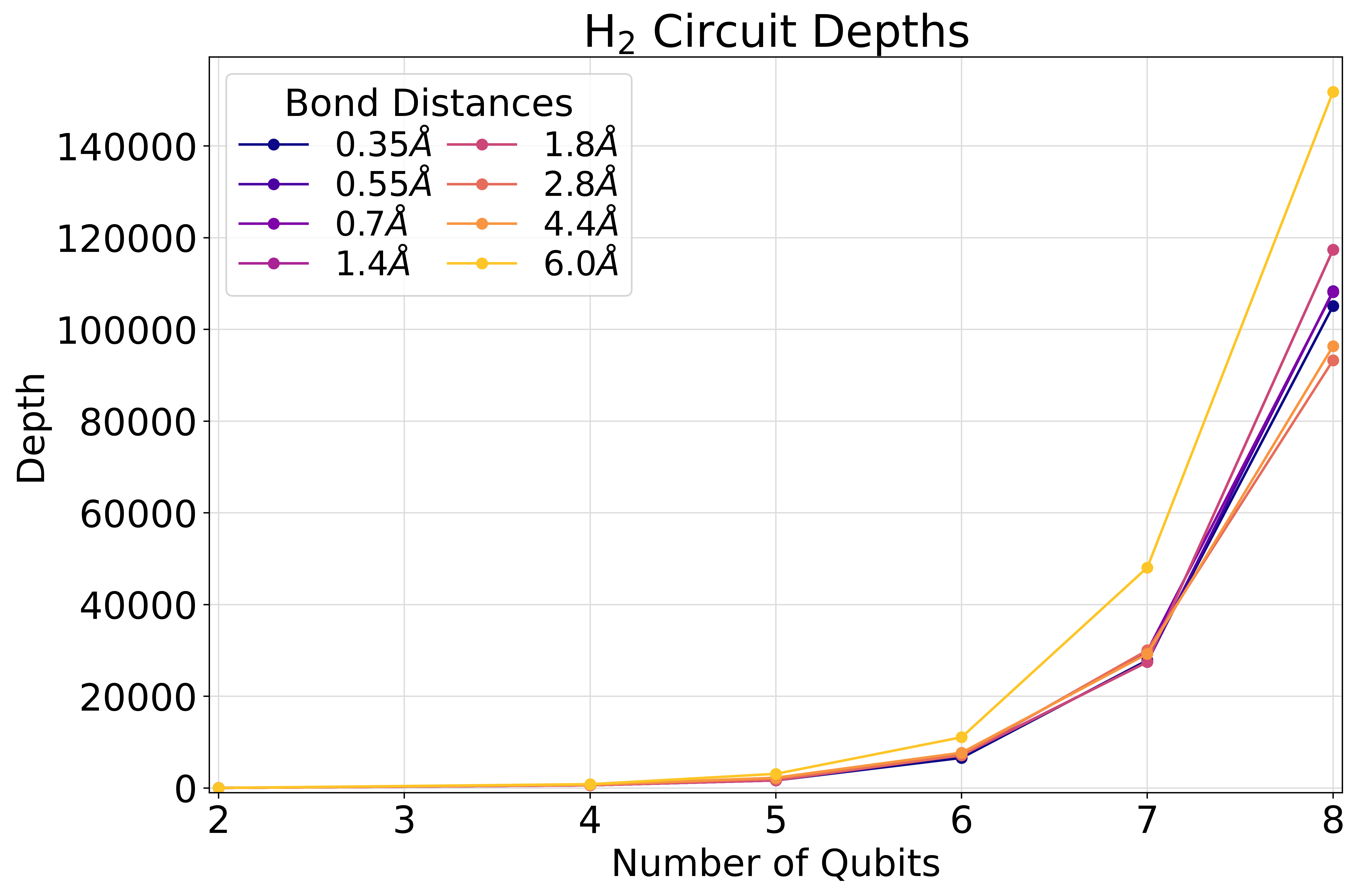}
    \caption{The circuit depths for our compact encoding as a function of the number of qubits for each bond distance.}
    \label{h2_depth}
\end{figure}

\begin{figure}[H]
    \centering
     \includegraphics[width=\linewidth]{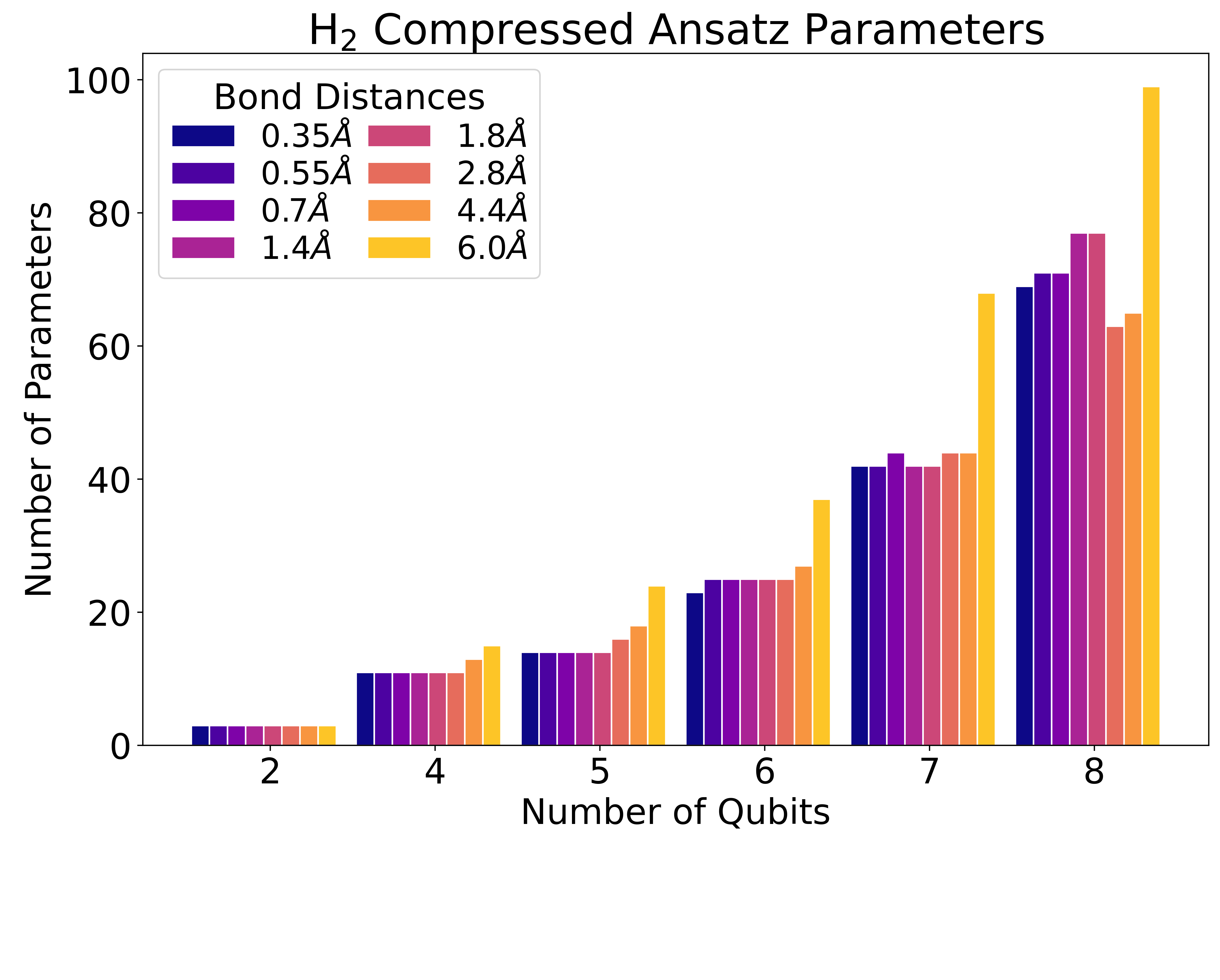}
     \caption{The number of parameters, including single excitations and perturbative double excitations, in the compact ansatz for each bond distance.}
     \label{h2_params} 
\end{figure}

 The biggest challenge facing compact fermion representations is the seemingly exponential growth of the circuit depth, as seen in Fig. \ref{h2_depth}. In our VQE implementation, we control the growth in depth by eliminating double excitations that do not contribute to the overall correlation energy using M{\o}ller-Plesset perturbation theory as found in Ref~\cite{Metcalf2020}. The correlation between circuit depth and the parameters is evident in Figs. \ref{h2_depth} and \ref{h2_params}. Fig. \ref{h2_params} illustrates the number of parameters that are in the VQE ansatz as a function of the number of qubits. There is significant growth in the number of parameters as the system size grows. Consequently, the compact mapping and ansatz presented in this paper are not scalable without implementing a method to reduce the circuit depth. 
 
  We can maximize compression on circuit depth using an optimization-based synthesis tool, LEAP, to synthesize a circuit with a minimal number of two qubit operations for our UCCSD matrix~\cite{LEAP2021}. Optimization-based synthesis tools suffer from scalability issues, so we test the tool on $\ce{H2}$ with 8 spin orbitals. We can achieve a circuit depth of 186 with 90 CNOT operations using this tool. Even with compression in qubit number and circuit depth, the energies obtained on publicly available quantum computing hardware are well outside of chemical accuracy. Extensive error mitigation techniques are required to extrapolate the desired wavefunction from the noisy result. 
  
  We compared an 8 spin orbital calculation by implementing VQE with a 4 qubit compact encoding and with a full 8 qubit using Jordan Wigner for $\ce{H2}$. In Table \ref{8spin_orbital}, we show the circuit depths prior to LEAP synthesis and the number of function calls to the classical optimizer. The number of parameters in the ansatz is equivalent for both, ranging from 11 to 15. Contrary to our expectations, the circuit depths using compact ansatz are not significantly larger than the depth using UCCSD. It is known that compact fermion representations have a larger gate overhead, however, we find in practice the difference is not all that significant. Interestingly, the gate depth for longer bond distance is less for the compact encoding, contrary to previous paper statements~\cite{moll2016optimizing}.
 
\begin{table}[]
{\renewcommand{\arraystretch}{1.17}
\resizebox{\columnwidth}{!}{%
\begin{tabular}{c@{\hspace{0.5cm}}c@{\hspace{0.5cm}}c@{\hspace{0.5cm}}c@{\hspace{0.5cm}}c}
\hline
\begin{tabular}[c]{@{}l@{}}Bond\\ Distances\end{tabular} & \begin{tabular}[c]{@{}l@{}l@{}}Full\\ Encoding\\ Depths\end{tabular} & \begin{tabular}[c]{@{}l@{}l@{}}Compact\\ Encoding\\ Depths\end{tabular} & \begin{tabular}[c]{@{}l@{}l@{}}Full\\ Encoding\\ Function Calls\end{tabular} &

\begin{tabular}[c]{@{}l@{}l@{}}Compact\\ Encoding\\ Function Calls\end{tabular} \\ \hline
0.35 \AA& 563& 600& 73& 57\\\arrayrulecolor{lightgray}\hline
0.55\AA& 563& 600& 77& 65\\\hline
0.7\AA& 563& 600& 97& 67\\\hline
1.4\AA& 563& 600& 388& 85\\\hline
1.8\AA& 563& 600& 667& 149\\\hline
2.8\AA& 563& 600& 643& 434\\\hline
4.4\AA& 739& 712& 676& 438\\\hline
6.0\AA& 979& 840& 796& 460\\ \arrayrulecolor{black}\hline
\end{tabular}}
}\caption{A benchmark for VQE circuit depths and function calls using 8 spin orbital $\ce{H2}$.}\label{8spin_orbital}
\end{table}

\begin{figure}[H]
    \centering
    {\includegraphics[width=\linewidth]{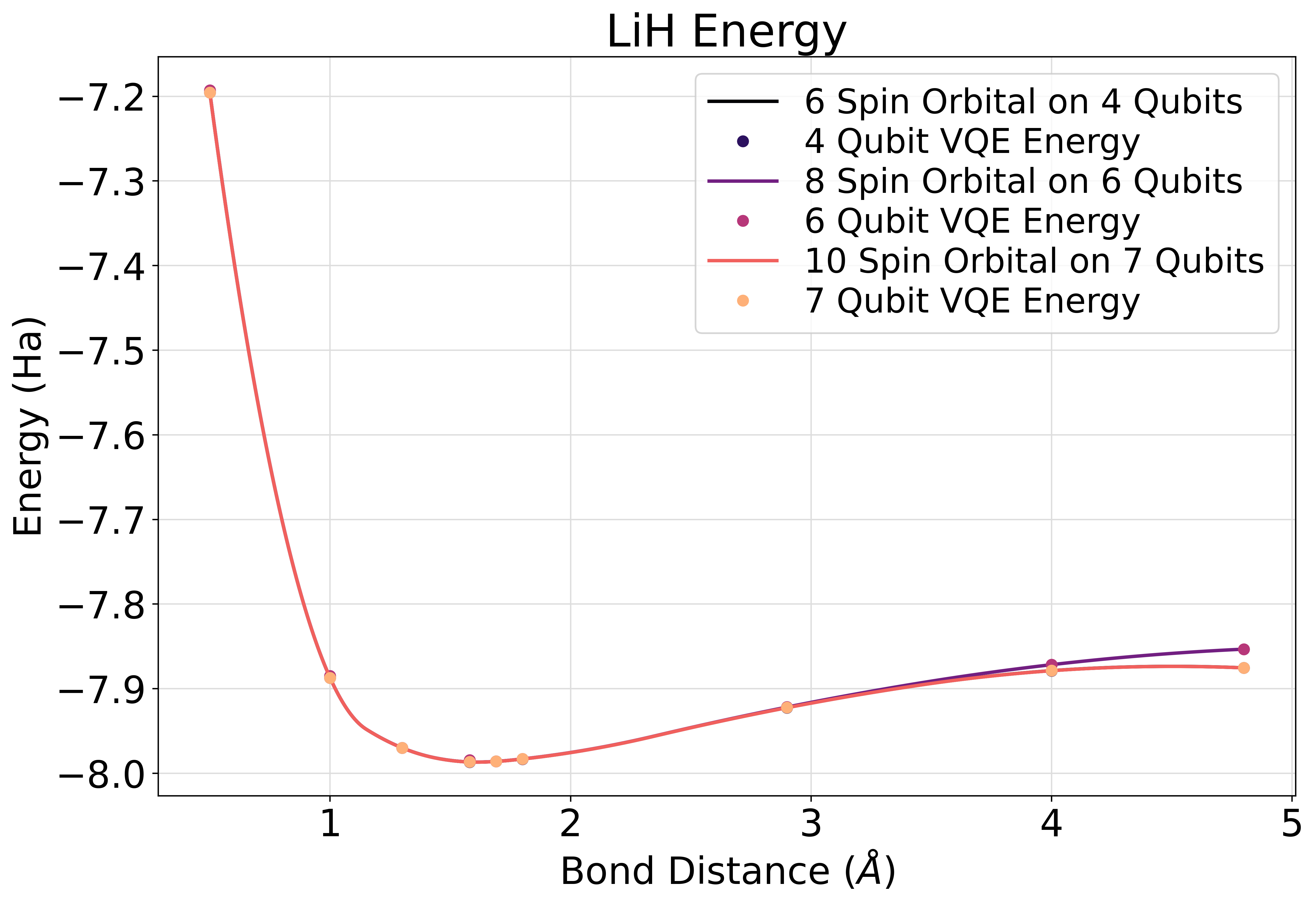}\label{lih}}
    \caption{The ground state energy as a function of the number of qubits using the compactly encoded Hamiltonian and Ansatz for $\ce{LiH}$ using 6, 8, and 10 active spin orbitals.}
    \label{Li_results}  
\end{figure}

A major bottleneck in VQE is the number of calls to the classical optimizer required to converge to the ground state energy. We find compact encoding converges to the correct energy with notably less function calls than the full encoding. This implies that state preparation by heuristic quantum methods is easier with compact encoding. It is difficult to evaluate the landscape of a high dimensional optimization space, so we conjecture the convergence results from differing landscapes with the two encodings. This is evident as the optimized values yielding the same ground state energy for the two cases differ. Comparable circuit depths and easier state preparation make a strong case for compact encodings.

Moreover, the mapping scheme and Ansatz we developed are generalized and can be applied to any molecule. 
We ran compact VQE for $\ce{LiH}$ with $N_\uparrow = 2$, $N_\downarrow = 2$, and $\{6, 8, 10\}$ active spin orbitals shown in Fig. \ref{Li_results}. Consider the case of 10 active spin orbitals. Here, $M=5$, thus, we can find the ground state energy using 7 qubits instead of 10. We calculate energies within chemical accuracy using compact VQE. The filling for $\ce{LiH}$ is near half with such a reduced active space, so other techniques to remove qubits are suitable. 

\section{Conclusion and Outlook}
In this work, we developed a new combinatoric compressed mapping, and a compressed UCCSD-inspired ansatz. The combinatoric compressed encoding maps fermions directly to the computational basis states rather than qubits. As a result, simulating the molecular Hamiltonian requires only $n = \ceil{\binom{M}{N_\uparrow}\binom{M}{N_\downarrow}}$ qubits. Moreover, we defined the cluster operator used in the compressed UCCSD-inspired ansatz such that it has the same dimensionally as the compactly encoded Hamiltonian. We assessed the performance of compact Hamiltonians and compact ansatz for variational quantum algorithms.

Classical composition of compact encoding schemes using full tomography scales exponentially with the number of qubits. A more scalable approach than we used in this paper is using sparse Hamiltonian simulation with oracular implementation of operators onto qubit basis states~\cite{berry2015hamiltonian, berry2017exponential}. Combinatoric ranking of the basis states compresses away the information needed to determine the basis state shift by fermionic operators, therefore, a look up table is required for this approach. Alternative compact mapping schemes that don't maximize compression simplify the oracle functions without the need for look up tables~\cite{Kirby_Compact_2021}.

Further, we argue that compact encodings are valuable in the NISQ era based on the results. The compact encoding we created significantly reduces the number of qubits required for simulation without increasing circuit depth. Additionally, we see a significant reduction in function calls made to the classical optimizer when running VQE. Combining our compact representation with techniques to downfold virtual integrals into reduced active spaces and sparsification of Hamiltonians will reduce the cost in qubits and gate depth in the near term~\cite{Metcalf2020,bauman2019downfolding, Dhawan2021}. 

\textbf{\textit{Acknowledgments - }}
This work was supported by the "Embedding QC into Many-body Frameworks for Strongly Correlated Molecular and Materials Systems” project, which is funded by the U.S. Department of Energy, Office of Science, Office of Basic Energy Sciences (BES), the Division of Chemical Sciences, Geosciences, and Biosciences. 
\bibliography{references.bib}

\appendix

\section{Classical Algorithm to Build Molecular Hamiltonian} \label{Sec:Hamiltonian_Alg}

Let $\mathcal{B}$ be the set of the basis elements of a given system consisting of $\ket{\sigma^\uparrow \sigma^\downarrow} = \ket{\sigma^\uparrow} \otimes \ket{\sigma^\downarrow}$ for $\ket{\sigma^\uparrow}\in S_\uparrow$ and  $\ket{\sigma^\downarrow} \in S_\downarrow$. Then, let $g$ be the function that maps the elements of the full basis $\ket{\sigma^\uparrow \sigma^\downarrow}$ to the reduced basis, $\mathcal{R}$. We will build the Hamiltonian using the elements of $\mathcal{R}$. Let's define the compact Hamiltonian, $\hat{H}_c$, as a sum of the one and two body operators as follows
\begin{align}
    \hat{H}_c = \hat{H_1} + \hat{H_2}
\end{align}
Where $\hat{H_1}$ and $\hat{H_2}$ are the one and two body operators in Eqn. \ref{Eq:MolHam}. To find $\hat{H}_1$ and $\hat{H}_2$ we must have the equivalent of $\hat{a}_i^\dagger \hat{a}_j$ and $\hat{a}_i^\dagger \hat{a}_j \hat{a}_k^\dagger \hat{a}_l$ acting in the reduced basis. Let's call $\hat{O}$ the operator that acts equivalently to $\hat{a}_i^\dagger \hat{a}_j$ in the reduced basis. For example, if $\hat{a}_i^\dagger \hat{a}_j$ maps $|\sigma_p^\uparrow \sigma_{p^\prime}^\downarrow\rangle \in \mathcal{B}$ to $\ket{\sigma_q^\uparrow \sigma_{q^\prime}^\downarrow} \in \mathcal{B}$,

\begin{align} \label{1bmap}
    \hat{a}_i^\dagger \hat{a}_j |\sigma_p^\uparrow \sigma_{p^\prime}^\downarrow\rangle = \ket{\sigma_q^\uparrow \sigma_{q^\prime}^\downarrow}
\end{align}
then $\hat{O}$ maps $\ket{I_p^{\uparrow \otimes \downarrow}} \in \mathcal{R}$ to $\ket{I_q^{\uparrow \otimes \downarrow}} \in \mathcal{R}$
\begin{align} \label{reduced_1bmap}
     \hat{O}\ket{I_p^{\uparrow \otimes \downarrow}} =\ket{I_q^{\uparrow \otimes \downarrow}}
\end{align}

We developed a method that produces the results given in eqns. \eqref{1bmap} and \eqref{reduced_1bmap}. We consider the subset of the full basis $\mathcal{B}$ that contains elements where $g\ket{\sigma^\uparrow \sigma^\downarrow} \in \mathcal{R}$. Then, after acting the creation and annihilation operators on those elements, we map them to the reduced basis. The details of the general algorithm to build the one and two-body matrices are given in algorithm~\ref{alg:cap}.

\begin{algorithm}
\caption{Building the compact Hamiltonian}
\label{alg:cap}
\begin{algorithmic}[1]
\State $n=\ceil{\log _{2}(\binom{M}{N_{\uparrow}}\binom{M}{N_{\downarrow}})}$
\State g: $\mathcal{B} \rightarrow \mathcal{R}$
\State Set $\hat{H}_1$ to a $2^n \times 2^n$ zero matrix \Comment{one-body}
\For{ all $\ket{\sigma_p^\uparrow \sigma_{p^\prime}^\downarrow} \in \mathcal{B}$ where $g\ket{\sigma_p^\uparrow \sigma_{p^\prime}^\downarrow} \in \mathcal{R}$} 
\State $\ket{\sigma_q^\uparrow \sigma_{q^\prime}^\downarrow} = \hat{a}_i^\dagger \hat{a}_j \ket{\sigma_p^\uparrow \sigma_{p^\prime}^\downarrow} $
\If{$g\ket{\sigma_q^\uparrow \sigma_{q^\prime}^\downarrow} \in \mathcal{R}$}
    \State $\ket{I_p^{\uparrow \otimes \downarrow}} = g(\ket{\sigma_p^\uparrow \sigma_{p^\prime}^\downarrow})$
    \State $\ket{I_q^{\uparrow \otimes \downarrow}} =g(\ket{\sigma_q^\uparrow \sigma_{q^\prime}^\downarrow})$
    \State $\hat{H_1}[\ket{I_p^{\uparrow \otimes \downarrow}}][\ket{I_q^{\uparrow \otimes \downarrow}}] \mathrel{{+}{=}} \hat{h}_1[i][j]$
    \State $\hat{H_1}[\ket{I_q^{\uparrow \otimes \downarrow}}][\ket{I_p^{\uparrow \otimes \downarrow}}] \mathrel{{+}{=}} \hat{h}_1^*[j][i]$ 
    
\EndIf
\EndFor
\State Set $\hat{H}_2$ to a $2^n \times 2^n$ zero matrix \Comment{two-body}
\For{ all $\ket{\sigma_p^\uparrow \sigma_{p^\prime}^\downarrow} \in \mathcal{B}$ where $g\ket{\sigma_p^\uparrow \sigma_{p^\prime}^\downarrow} \in \mathcal{R}$} 
\State $\ket{\sigma_q^\uparrow \sigma_{q^\prime}^\downarrow} = \hat{a}_i^\dagger \hat{a}_j \ket{\sigma_p^\uparrow \sigma_{p^\prime}^\downarrow}$
\If{$g\ket{\sigma_q^\uparrow \sigma_{q^\prime}^\downarrow} \in \mathcal{R}$}
    \State $\ket{\sigma_t^\uparrow \sigma_{t^\prime}^\downarrow} = \hat{a}_k^\dagger \hat{a}_l \ket{\sigma_q^\uparrow \sigma_{q^\prime}^\downarrow}$
    \If{$f\ket{\sigma_t^\uparrow \sigma_{t^\prime}^\downarrow} \in \mathcal{R}$}
        \State $\ket{I_p^{\uparrow \otimes \downarrow}} = g(\ket{\sigma_p^\uparrow \sigma_{p^\prime}^\downarrow})$
        \State $\ket{I_t^{\uparrow \otimes \downarrow}} = g(\ket{\sigma_t^\uparrow \sigma_{t^\prime}^\downarrow})$
        \State $\hat{H_2}[\ket{I_p^{\uparrow \otimes \downarrow}}][\ket{I_t^{\uparrow \otimes \downarrow}}] \mathrel{{+}{=}} \hat{h}_2[i][j][k][l]$
        \State $\hat{H_2}[\ket{I_t^{\uparrow \otimes \downarrow}}][\ket{I_p^{\uparrow \otimes \downarrow}}] \mathrel{{+}{=}} \hat{h}^*_2[k][l][i][j]$ 
    \EndIf
\EndIf
    \If{$k = j$}
        \State $\ket{\sigma_q^\uparrow \sigma_{q^\prime}^\downarrow} = \hat{a}_i^\dagger \hat{a}_l \ket{\sigma_p^\uparrow \sigma_{p^\prime}^\downarrow}$
        \If{$g\ket{\sigma_q^\uparrow \sigma_{q^\prime}^\downarrow} \in \mathcal{R}$}
            \State $\ket{I_q^{\uparrow \otimes \downarrow}} = g(\ket{\sigma_q^\uparrow \sigma_{q^\prime}^\downarrow})$
            \State $\ket{I_p^{\uparrow \otimes \downarrow}} = g(\ket{\sigma_p^\uparrow \sigma_{p^\prime}^\downarrow})$
            \State $\hat{H_2}[\ket{I_p^{\uparrow \otimes \downarrow}}][\ket{I_q^{\uparrow \otimes \downarrow}}] \mathrel{{-}{=}} \hat{h}_2[i][j][k][l]$
            \State $\hat{H_2}[\ket{I_q^{\uparrow \otimes \downarrow}}][\ket{I_p^{\uparrow \otimes \downarrow}}] \mathrel{{-}{=}} \hat{h}_2^*[k][l][i][j]$
        \EndIf
    \EndIf
\EndFor
\State $\hat{H}_c = \hat{H}_1 + \hat{H}_2$
\end{algorithmic}
\end{algorithm}

\end{document}